\documentclass{article}

\usepackage{mathrsfs}
\usepackage{amsfonts}
\usepackage{amsmath}
\usepackage{amssymb}
\usepackage{graphicx}

\title{Possible physical universes}
\author{Gordon McCabe}

\begin{document}

\maketitle

\begin{abstract}

The purpose of this paper is to discuss the various types of
physical universe which could exist according to modern
mathematical physics. The paper begins with an introduction that
approaches the question from the viewpoint of ontic structural
realism. Section 2 takes the case of the `multiverse' of spatially
homogeneous universes, and analyses the famous Collins-Hawking
argument, which purports to show that our own universe is a very
special member of this collection. Section 3 considers the
multiverse of all solutions to the Einstein field equations, and
continues the discussion of whether the notions of special and
typical can be defined within such a collection.

\end{abstract}

\section{Introduction}

Einstein famously claimed that ``what really interests me is
whether God had any choice in the creation of the world." This is
generally considered to be a whimsical version of the question,
`is there only one logically possible physical universe?' The
modern answer to this question is: `apparently not!'

In this paper, we will approach the notion of possible physical
universes using the philosophical doctrine of structural realism,
which asserts that, in mathematical physics at least, the physical
domain of a true theory is an instance of a mathematical
structure.\footnote{This notion was originally advocated by
Patrick Suppes (1969), Joseph Sneed (1971), Frederick Suppe
(1989), and others.} It follows that if the domain of a true
theory extends to the entire physical universe, then the entire
universe is an instance of a mathematical structure. Equivalently,
it is asserted that the physical universe is isomorphic to a
mathematical structure. Let us refer to this proposal as
`universal structural realism'.

Whilst the definition of structural realism is most frequently
expressed in terms of the set-theoretical, Bourbaki notion of a
species of mathematical structure, one can reformulate the
definition in terms of other approaches to the foundations of
mathematics, such as mathematical category theory. In the latter
case, one would assert that our physical universe is an object in
a mathematical category.

Many of the authors writing about such ideas neglect to assert
that the universe is \emph{an instance} of a mathematical
structure, but instead claim that the universe \emph{is} a
mathematical structure. In some cases this can be taken as merely
an abbreviated form of speech; in others, the distinction in
meaning is deliberate, and such authors may be sympathetic to the
notion that the physical universe is nothing but form, and the
notion of substance has no meaning. This corresponds to Ladyman's
distinction (1998) between the ontic and epistemic versions of
structural realism. Whilst the epistemic version accepts that
mathematical physics captures the structure possessed by the
physical world, it holds that there is more to the physical world
beyond the structure that is possesses. In contrast, the ontic
version holds that the structure of the physical world is the only
thing which exists.

Those expressions of universal structural realism which state that
`the' physical universe is an instance of a mathematical
structure, tacitly assume that our physical universe is the only
physical universe. If one removes this assumption, then universal
structural realism can be taken as the two-fold claim that (i) our
physical universe is an instance of a mathematical structure, and
(ii), other physical universes, if they exist, are either
different instances of the same mathematical structure, or
instances of different mathematical structures. Given that
mathematical structures are arranged in tree-like hierarchies,
other physical universes may be instances of mathematical
structures which are sibling to the structure possessed by our
universe. In other words, the mathematical structures possessed by
other physical universes may all share a common parent structure,
from which they are derived by virtue of satisfying additional
conditions. This would enable us to infer the mathematical
structure of other physical universes by first generalizing from
the mathematical structure of our own, and then classifying all
the possible specializations of the common, generic structure.

It is common these days to refer to the hypothetical collection of
all physical universes as the multiverse. I will refrain from
using the phrase `ensemble of all universes', because an ensemble
is typically considered to be a space which possesses a
probability measure, and it is debatable whether the universe
collections considered in this paper possess a natural probability
measure. It should also be emphasised that no physical process
will be suggested to account for the existence of the multiverses
considered in this paper. This paper will not address those
theories, such as Linde's chaotic inflation theory (1983a and
1983b) or Smolin's theory of cosmological natural selection
(1997), which propose physical processes that yield collections of
universes or universe-domains. The universes considered in this
paper are mutually disjoint, and are not assumed to be the outcome
of a common process, or the outcome of any process at all.

If a physical universe is conceived to be an instance of a
mathematical structure, i.e. a structured set, then it is natural
to suggest that the multiverse is a collection of such instances,
or a collection of mathematical structures. Tegmark, for example,
characterises this view as suggesting that ``some subset of all
mathematical structures...is endowed with...physical existence,"
(1998, p1). As Tegmark points out, such a view of the multiverse
fails to explain why some particular collection of mathematical
structures is endowed with physical existence rather than another.
His response is to suggest that all mathematical structures have
physical existence.

Like many authors, Tegmark implicitly assumes that the physical
universe must be a structured set (or an instance thereof). As
Rucker asserts, ``if reality is physics, if physics is
mathematics, and if mathematics is set theory, then everything is
a set," (Rucker 1982, p200). However, as already alluded to,
mathematics cannot be identified with set theory, hence it doesn't
follow that everything is a set. There are mathematical objects,
such as topoi, which are not sets. Hence, it may be that the
physical universe is not a set.

Category theory is able to embrace objects which are not sets. A
category consists of a collection of objects such that any pair of
objects has a collection of morphisms between them. The morphisms
satisfy a binary operation called composition, which is
associative, and each object has a morphism onto itself called the
identity morphism. For example, the category \textbf{Set} contains
all sets as objects and the functions between sets as morphisms;
the category of topological spaces has continuous functions as
morphisms; and the category of smooth manifolds has smooth
(infinitely-differentiable) maps as morphisms. One also has
categories such as the category \textbf{Top} of all topoi, in
which the morphisms need not be special types of functions and the
objects need not be special types of set. One can therefore
generalize the central proposition of universal structural
realism, to assert that the physical universe is an object in a
mathematical category.

If the physical universe is not a structured set of some kind,
then the multiverse may not be a collection of structured sets
either. In fact, in the case of Tegmark's suggestion that the
multiverse consists of all mathematical sets, it is well-known
that there is no such thing as the set of all sets, so Tegmark
would be forced into conceding that the multiverse is not a set
itself, but the category of all sets, perhaps. More generally, one
could suggest that the multiverse is a collection of mathematical
objects, which may or may not be structured sets. The multiverse
may be a category, but it may not be a category of structured
sets. The analogue of Tegmark's suggestion here would perhaps be
to propose that all categories physically exist. Categories can be
related by maps called `functors', which map the objects in one
category to the objects in another, and which map the morphisms in
one category to the morphisms in the other category, in a way
which preserves the composition of morphisms. Furthermore, one can
relate one functor to another by something called a `natural
transformation'. In effect, one treats functors as higher-level
objects, and natural transformations are higher-level morphisms
between these higher-level objects. Accordingly, natural
transformations are referred to as `2-morphisms'. Whilst a
category just possesses objects and morphisms, a 2-category
possesses objects, morphisms and 2-morphisms. The collection of
all categories is a 2-category in the sense that it contains
categories, functors between categories, and natural
transformations between functors; each category is an object, the
functors between categories are morphisms, and the natural
transformations are 2-morphisms. However, the 2-category of all
categories is only one example of a 2-category. A 2-category which
has categories as objects, need not contain all categories, and a
2-category need not even have categories as objects. To fully
develop Tegmark's suggestion, one would need to propose that all
2-categories physically exist, and the latter itself is just one
example of a 3-category. One would need to continue indefinitely,
with the category of all $n$-categories always just one example of
an $n+1$-category, for all $n \in \mathbb{N}$.

Let us return, however, to the notion that a physical universe is
an instance of a structured set, and let us refine this notion in
terms of mathematical logic. A theory is a set of sentences, in
some language, which is closed under logical implication. A model
for a set of sentences is an interpretation of the langauge in
which those sentences are expressed, which renders each sentence
as true. Each theory in mathematical physics has a class of models
associated with it. As Earman puts it, ``a practitioner of
mathematical physics is concerned with a certain mathematical
structure and an associated set $\mathfrak{M}$ of models with this
structure. The...laws $L$ of physics pick out a distinguished
sub-class of models $\mathfrak{M}_L := \text{Mod}(L) \subset
\mathfrak{M}$, the models satisfying the laws $L$ (or in more
colorful, if misleading, language, the models that ``obey" the
laws $L$)," (p4, 2002). Hence, any theory whose domain extends to
the entire universe, (i.e. any cosmological theory), has a
multiverse associated with it: namely, the class of all models of
that theory. For a set $\Sigma$ of sentences, let $\text{Mod} \;
\Sigma$ denote the class of all models of $\Sigma$. The class
$\text{Mod} \; \Sigma$, if non-empty, is too large to be a set,
(Enderton 2001, p92). Hence, in this sense, the multiverse
associated with any cosmological theory is too large to be a set
itself. At face value, this seems to contradict the fact that,
whilst many of the multiverses considered in cosmology possess an
infinite cardinality, and may even form an infinite-dimensional
space, they are, nevertheless, sets. This apparent contradiction
arises because physicists tacitly restrict the range of
interpretations of the languages in which their theories are
expressed, holding the meaning of the predicates fixed, but
allowing the variables to range over different domains.

A theory $T$ is complete if for any sentence $\sigma$, either
$\sigma$ or $\neg \sigma$ belongs to $T$. If a sentence is true in
some models of a theory but not in others, then the theory is
incomplete. Godel's first incompleteness theorem demonstrates that
Peano arithmetic is incomplete. Hence, any theory which includes
Peano arithmetic will also be incomplete. If a final theory of
everything includes Peano arithmetic, (an apparently moderate
requirement), then the final theory will be incomplete. Such a
final theory of everything would not eliminate contingency. There
would be sentences true in some models of a final theory, but not
true in others. Hence, the multiverse hypothesis looms over even a
hypothetical final theory of everything. Godel's incompleteness
theorem dispels the possibility that there is only one logically
possible physical universe.

Jesus Mosterin (2004) points out that ``the set of all possible
worlds is not at all defined with independence from our conceptual
schemes and models. If we keep a certain model (with its
underlying theories and mathematics) fixed, the set of the
combinations of admissible values for its free parameters gives us
the set of all possible worlds (relative to that model). It
changes every time we introduce a new cosmological model (and we
are introducing them all the time). Of course, one could propose
considering the set of all possible worlds relative to all
possible models formulated in all possible languages on the basis
of all possible mathematics and all possible underlying theories,
but such consideration would produce more dizziness than
enlightment."

Mosterin's point here is aimed at the anthropic principle, and the
suggestion that there are multiverses which realize all possible
combinations of values for the free parameters in physical
theories such as the standard model of particle physics. At face
value, these are different types of multiverse than the ones
proposed in this paper, which are obtained by varying mathematical
structures, and by taking all the models of a fixed mathematical
structure, rather than by taking all the values of the free
parameters within a theory. However, the values chosen for the
free parameters of a theory correspond to a choice of model in
various parts of the theory. For example, the free parameters of
the standard model of particle physics include the coupling
constants of the strong and electromagnetic forces, two parameters
which determine the Higgs field potential, the Weinberg angle, the
masses of the elementary quarks and leptons, and the values of
four parameters in the Kobayashi-Maskawa matrix which specifies
the `mixing' of the $\{d,s,b\}$ quark flavours in weak force
interactions. The value chosen for the coupling constant of a
gauge field with gauge group $G$ corresponds to a choice of metric
in the lie algebra $\mathfrak{g}$, (Derdzinksi 1992, p114-115);
the Weinberg angle corresponds to a choice of metric in the lie
algebra of the electroweak force, (ibid., p104-111); the values
chosen for the masses of the elementary quarks and leptons
correspond to the choice of a finite family of irreducible unitary
representations of the local space-time symmetry group, from a
continuous infinity of alternatives on offer; and the choice of a
specific Kobayashi-Maskawa matrix corresponds to the selection of
a specific orthogonal decomposition $\sigma_{d'} \oplus
\sigma_{s'} \oplus \sigma_{b'}$ of the fibre bundle which
represents a generalization of the $\{d,s,b\}$ quark flavours,
(ibid., p160).

In general relativity, a universe is represented by a
4-dimensional differential manifold $\mathcal{M}$ equipped with a
metric tensor field $g$ and a set of matter fields and gauge force
fields $\{\phi_i \}$ which generate an energy-stress-momentum
tensor $T$ that satisfies the Einstein field equations
$$ T = 1/(8\pi G)(\text{Ric} - 1/2 \; \text{S} \, g) \,.
$$ $Ric$ denotes the Ricci tensor field determined by $g$,
and $\text{S}$ denotes the curvature scalar field. The matter
fields have distinctive equations of state, and include fluids,
scalar fields, tensor fields, and spinor fields. Gauge force
fields, such as electromagnetism, are described by $n$-form
fields. Hence, one can define a general relativistic multiverse to
be the class of all models of such $n$-tuples
$\{\mathcal{M},g,\phi_1,...\}$, interpreted in this restricted
sense.

Alternatively, to take an example suggested by David Wallace
(2001), quantum field theory represents a universe to be a
Lorentzian manifold $(\mathcal{M},g)$ which is equipped with a
Hilbert space $\mathscr{H}$, a density operator $\rho$ on
$\mathscr{H}$, and a collection of operator-valued distributions
$\{\hat{\phi}_i \}$ on $\mathcal{M}$ which take their values as
bounded self-adjoint operators on $\mathscr{H}$. A quantum field
theory multiverse is the class of all models of such $n$-tuples
$\{\mathcal{M},g,\mathscr{H},\rho, \hat{\phi}_1,...\ \}$,
interpreted in this restricted sense.

Assuming that cosmological theories are
axiomatizable,\footnote{i.e assuming that there is a decidable set
of sentences (Enderton p62) in such a theory, from which all the
sentences of the theory are logically implied.} one can abstract
from the theories which may just apply to our universe or a
collection of universes in a neighbourhood of our own, by varying
or relaxing the axioms. Each set of axioms has its own class of
models, which, in this context, provides a multiverse. Hence, by
varying or relaxing the axioms of a theory which is empirically
verified in our universe, one generates a tree-like hierarchy of
multiverses, some of which are sibling to the original class of
models, and some of which are parents or `ancestors' to the
original class.

If an empirically verified theory, with a class of models
$\mathfrak{M}$, and a set of laws $L$, defines a subclass of
models $\mathfrak{M}_L$, and if those laws contain a set of free
parameters $\{p_i : i = 1,...,n\}$, then one has a different class
of models $\mathfrak{M}_{L(p_{i})}$ for each set of combined
values of the parameters $\{p_i\}$. These multiverses are sibling
to each other in the hierarchy. By relaxing the requirement that
any set of laws be satisfied in the mathematical structure in
question, one obtains a multiverse $\mathfrak{M}$ which is parent
to all the multiverses $\mathfrak{M}_{L(p_{i})}$. By varying the
axioms that define the original mathematical structure, one
obtains sibling mathematical structures which have classes of
model $\{\mathfrak{N}_j \}$ sibling to $\mathfrak{M}$. By relaxing
the axioms that define the original mathematical structure, one
steadily obtains more general mathematical structures which have
more general classes of model $\mathfrak{P}_k \supset
\mathfrak{M}$. Each such $\mathfrak{P}_k$ is a parent or ancestor
in the hierarchy to all the multiverses $\{\mathfrak{N}_j \}$. For
example, if we take general relativistic cosmology to provide a
theory of our own universe, one can obtain a selective hierarchy
of multiverses such as the following:

\begin{itemize}
\item{All Friedmann-Robertson-Walker (FRW) spatially isotropic Lorentzian 4-manifolds and matter field pairings.}
\item{All Bianchi spatially homogeneous Lorentzian 4-manifolds and matter field pairings.}
\item{All Lorentzian 4-manifolds which solve the canonical/initial-value formulation of the Einstein field equations
with respect to some combination of matter fields and gauge
fields,\footnote{Note that matter fields and gauge force fields
must satisfy their own constraint equations and evolution
equations.} for a fixed spatial topology $\Sigma$.}
\item{All Lorentzian 4-manifolds which satisfy the Einstein field equations with respect to some combination of matter
fields and gauge fields, for the value of the gravitational
constant $G \approx 6.67 \times 10^{-8}cm^3 g^{-1} s^{-2}$ that we
observe in our universe.}
\item{All Lorentzian 4-manifolds which satisfy the Einstein field equations with respect to some combination of
matter fields and gauge fields, for a different value of the
gravitational constant $G$.}
\item{All Lorentzian 4-manifolds equipped with some combination of matter fields and gauge fields, irrespective
of whether they satisfy the Einstein field equations.}
\item{All Lorentzian 4-manifolds.}
\item{All manifolds of arbitrary dimension and geometrical signature, which satisfy the Einstein field equations
with respect to some combination of matter fields and gauge
fields.}
\item{All 4-manifolds equipped with combinations of smooth tensor and spinor
fields\footnote{Note that \emph{global} spinor fields require a manifold
to possess a spin structure.} which satisfy differential
equations.}
\item{All 4-manifolds.}
\item{All differential manifolds of any dimension equipped with combinations of smooth tensor and
spinor fields which satisfy differential equations.}
\item{All differential manifolds.}
\item{All topological manifolds equipped with combinations of continuous functions which satisfy algebraic\footnote
{Algebraic equations involve only the operations defined upon the
algebra of functions $\mathscr{A}$. This means operations such as
scalar multiplication, sum, product, and derivation. In this
context, a derivation of $\mathscr{A}$ is a mapping $X :
\mathscr{A} \rightarrow \mathscr{A}$ which is linear, $X(af + bg)
= aXf + bXg$, and which satisfies the so-called Leibniz rule,
$X(fg) = fX(g) + X(f)g$.} equations.}
\item{All topological spaces with the cardinality of the continuum equipped with combinations of continuous functions
which satisfy algebraic equations.}
\item{All topological spaces of any cardinality equipped with combinations of continuous functions
which satisfy algebraic equations.}
\item{All topological spaces.}
\item{All sets equipped with combinations of functions which satisfy algebraic equations.}
\item{All sets.}
\item{All categories.}
\end{itemize}

Mosterin comments acerbically that ``authors fond of many
universes talk about them in a variety of incompatible ways. The
totality of the many universes accepted by an author forms the
multiverse for that author. There are at least as many multiverses
as authors talking about them; in fact, there are more, as some
authors have several multiverses to offer," (2004). Bearing this
in mind, it should be declared that this paper is only interested
in multiverses consisting of the models of empirically verified
theories, or generalisations obtained from such theories. Hence,
multiverses derived from supersymmetry, supergravity, superstring
or M theory, will not be considered.

From the list of multiverses above, we now proceed to consider a
couple of interesting cases. In particular, we will be interested
in understanding how special or typical our own universe is with
respect to these multiverses.

\section{The multiverse of spatially homogeneous models}

The purpose of this section is to analyse an argument by Collins
and Hawking that our own spatially isotropic universe is extremely
atypical even in the space of spatially homogeneous models. To
gain some understanding of the argument, however, we need to begin
with some facts about spatially homogeneous models.

It is generally believed that, up to local isometry, a spatially
homogeneous model, equipped with a fluid satisfying a specific
equation of state, can be uniquely identified by specifying both
its Bianchi type, and by specifying its dynamical history. The
Bianchi type classifies the 3-dimensional Lie algebra of Killing
vector fields on the spatially homogeneous hypersurfaces of such a
model.

In any 3-dimensional Lie algebra, the space of all possible bases
is 9-dimensional, and the general linear group $GL(3,\mathbb{R})$
acts simply transitively upon this space of bases. Hence, if one
fixes a basis, then one can establish a one-to-one mapping between
the bases in the Lie algebra and the matrices in
$GL(3,\mathbb{R})$. Needless to say, $GL(3,\mathbb{R})$ is a
9-dimensional group. Now, specifying the structure constants of a
3-dimensional Lie algebra relative to a particular basis uniquely
identifies a particular Bianchi type. Given a Bianchi type fixed
in such a manner, one can allow $GL(3,\mathbb{R})$ to act upon the
space of bases in the Lie algebra. Under some changes of basis,
the structure constants will change, whilst under other changes of
basis, they will remain unchanged. Hence, the action of
$GL(3,\mathbb{R})$ upon the space of structure constants for a
particular Bianchi type is multiply transitive; the action of
$GL(3,\mathbb{R})$ upon the space of structure constants has a
non-trivial stability subgroup. The dimension of this stability
subgroup depends upon the Bianchi type.

Another way of looking at this is to consider the 6-dimensional
space of all possible structure constants for all possible
3-dimensional Lie algebras. The general linear group
$GL(3,\mathbb{R})$ acts upon this space. The
$GL(3,\mathbb{R})$-action does not map the structure constants of
one Bianchi type into another; changing the basis in a Lie algebra
will not give you the structure constants of a different Lie
algebra. Hence, each orbit of $GL(3,\mathbb{R})$ in the space of
all structure constants corresponds to a particular Bianchi type.
The dimension $p$ of each orbit is given in the table below,
(Collins and Hawking 1973, p321; Hewitt \emph{et al} 1997, p210;
MacCallum 1979, p541). The dimension of the
$GL(3,\mathbb{R})$-orbit for each Bianchi type equals the number
of free parameters required to specify the structure constants for
that Bianchi type. Given that $GL(3,\mathbb{R})$ is a
9-dimensional group, it follows that its stability group at each
point in the space of structure constants will be $9-p$
dimensional. This is the dimension of the space of bases under
which the structure constants are unchanged.

\begin{table}[h]
\caption{Dimension of the $GL(3,\mathbb{R})$-orbits for each
Bianchi type.} \label{Bianchib}
\begin{center}
\begin{tabular}{|c|c|}
  \noalign{\hrule}
Type & Dimension \\ \hline
  $\textrm{I}$ & 0 \\
  $\textrm{II}$ & 3 \\
$\textrm{VI}_0$ & 5 \\
$\textrm{VII}_0$ & 5 \\
$\textrm{VIII}$ & 6 \\
 $\textrm{IX}$ & 6 \\
  $\textrm{V}$ & 3 \\
 $\textrm{IV}$ & 5 \\
$\textrm{VI}_h$ & 6 \\
 $\textrm{VII}_h$ & 6 \\ \hline
\end{tabular}
\end{center}
\end{table}

Collins and Hawking (1973) argued that the set of spatially
homogeneous but anisotropic models which tend towards isotropy as
$t \rightarrow \infty$ is of measure zero in the set of all
spatially homogeneous initial data. They first excluded all the
Bianchi types whose metrics are of measure zero in the space of
all 3-dimensional homogeneous metrics. This includes all the
Bianchi types whose $GL(3,\mathbb{R})$-orbits are of dimension
less than 6. They then excluded types $\textrm{VI}_h$ and
$\textrm{VIII}$ on the grounds that they do not contain any FRW
models as limiting cases. However, the class of Bianchi type
$\textrm{VII}_h$ models contain the ever-expanding FRW universes,
with spatial curvature $k<0$, as special cases, and the class of
type $\textrm{IX}$ models contain the closed $k>1$ FRW universes.
In the class of type $\textrm{VII}_h$ models, where $\sigma_{ij}$
is the shear tensor and $H$ is a generalized Hubble parameter,
approach to isotropy was defined to mean that the `distortion'
$\sigma/H \rightarrow 0$ as $t \rightarrow \infty$, and that the
cumulative distortion $\int^t \sigma \;dt$ approaches a constant
as $t \rightarrow \infty$, where $\sigma = {\sigma^i}_i$. Collins
and Hawking concluded that there is no open neighbourhood of the
FRW models in the space of either type $\textrm{VII}_h$ or type
$\textrm{IX}$ metrics which tends towards isotropy. However, they
did find that in the space of type $\textrm{VII}_0$ metrics, which
contains the $k=0$ FRW universes as special cases, and with the
matter assumed to be zero-pressure `dust', there is an open
neighbourhood about such FRW universes which do approach isotropy.
However, the space of type $\textrm{VII}_0$ metrics is of measure
zero in the space of all homogeneous metrics. Moreover, the
assumption of a zero-pressure matter field prevents one applying
this result to universes which have a radiated-dominated phase,
such as our own is believed to have undergone in its early
history.

From their conclusion that the set of spatially homogeneous models
which approach isotropy as $t \rightarrow \infty$ is of measure
zero, Collins and Hawking also inferred that isotropic models are
unstable under spatially homogeneous perturbations. In other
words, it was argued that a model which is initially almost
isotropic and spatially homogeneous, will tend towards anisotropy.
However, Barrow and Tipler correctly point out that the
requirement $\sigma/H \rightarrow 0$ is the condition of
`asymptotic stability', and the open FRW universe in the type
$\textrm{VII}_h$ Bianchi class is stable under spatially
homogeneous perturbations in the sense that $\sigma/H$ approaches
a constant. According to Barrow and Tipler this shows ``that
isotropic open universes are stable in the same sense that our
solar system is stable. As $t \rightarrow \infty$ there exist
spatially homogeneous perturbations with $\sigma/H \rightarrow
\text{constant}$ but there are none with $\sigma/H \rightarrow
\infty$. The demand for asymptotic stability is too strong a
requirement," (1986, p425). Whilst the set of spatially
homogeneous models which approach isotropy is of measure zero, our
own universe is not exactly isotropic, it is merely
`almost-isotropic', and, moreover, it is growing increasingly
anisotropic as a function of time. Collins and Hawking fail to
demonstrate that almost-isotropic spatially homogeneous models are
of measure zero in the space of spatially homogeneous models.
Moreover, our own universe is almost a FRW model in the sense that
it has approximate spatial homogeneity and approximate spatial
isotropy. Our own universe, then, has been perturbed by
inhomogeneous perturbations. Collins and Hawking fail to
demonstrate that almost-isotropic and almost-homogeneous
universes, like our own, are of measure zero in the space of
almost homogeneous models.

To derive their conclusion that type $\textrm{VII}_h$ models fail
to isotropize, Collins and Hawking assumed various reasonable
conditions defining the nature of matter, but they also assumed a
zero cosmological constant, an assumption which recent
astronomical evidence for an accelerating universe has cast into
serious doubt. Wald (1983) demonstrated that all initially
expanding non-type $\textrm{IX}$ Bianchi models with spatial
curvature $k \leq 0$ and a positive cosmological constant, do in
fact isotropize, expanding exponentially, and tending toward de
Sitter space-time, with $\sigma/H \rightarrow 0$ as $t \rightarrow
\infty$. However, Wald's result only applies to Bianchi type
$\textrm{IX}$ models if one assumes that the cosmological constant
is initially large in comparison with the scalar curvature of the
hypersurfaces of homogeneity, (Earman and Mosterin, 1999). In
addition, Wald's result has not been extended to spatially
inhomogeneous universes. The attempt by Jensen and Stein-Schabes
(1987) to extend Wald's result makes the physically unacceptable
assumption that the scalar curvature is non-positive throughout
the space-time. Even with initial non-positive scalar curvature,
in a spatially inhomogeneous universe one would expect pockets of
positive curvature to form with the passage of time, such as those
around black holes, stars and planets. The exterior Schwarzschild
solution, generalized to the case of a positive cosmological
constant, has such positive curvature, and doesn't evolve towards
de Sitter space-time, (Earman and Mosterin 1999). It remains an
open question whether initially expanding, spatially inhomogeneous
universes of initial non-positive curvature, which develop
localized pockets of positive curvature, evolve towards de Sitter
space-time if they have a positive cosmological constant.

It is true, however, that the finite-dimensional space of exact
FRW models is of measure zero in the finite-dimensional space of
spatially homogeneous models, and Collins and Hawking demonstrate
that spatially homogeneous anisotropic models do not generally
approach isotropic ones, unless the cosmological constant is
non-zero. As we will see, the space of spatially homogeneous
models is itself the complement of an open dense subset in the
infinite-dimensional space of all solutions to the Einstein field
equations. Hence, the space of exact FRW models is as numerous in
the space of spatially homogeneous models as the integers are in
the set of real numbers, and the set of spatially homogeneous
models is as prevalent in the space of all solutions to the
Einstein field equations as the points in the surface of solid
ball are to the points in the interior of the solid ball.

\section{The multiverse of all solutions to the Einstein field equations}

For each 4-dimensional manifold $\mathcal{M}$, one has the space
of all solutions to the Einstein field equations on that manifold,
$\mathscr{E}(\mathcal{M})$. Rather than dealing with matter field
solutions, existing analysis has concentrated on vacuum solutions
of the Einstein field equations, and yet further restrictions have
been placed on the topology and geometry of the solutions
considered. For example, work has been done on the space
$\tilde{\mathscr{E}}(\mathcal{M})$ of vacuum solutions of a
globally hyperbolic space-time with compact spatial topology
$\Sigma$, containing a constant mean extrinsic curvature Cauchy
hypersurface. The restrictions placed upon such solution spaces
means that they really just contain the solutions of initial data
sets. Nothing is revealed about spaces of solutions which do not
admit an initial-value formulation.

$\tilde{\mathscr{E}}(\mathcal{M})$ is not a manifold, but a
stratified space. The points representing solutions with a
non-trivial isometry group are said to have conical
neighbourhoods. The strata consist of space-times with conjugate
isometry groups, (Isenberg and Marsden 1982, p187).

Two space-time solutions are considered to be physically
equivalent if one is isometric to the other, hence work has been
done on the quotient space
$\tilde{\mathscr{E}}(\mathcal{M})/\mathscr{D}(\mathcal{M})$ with
respect to the diffeomorphism group $\mathscr{D}(\mathcal{M})$ of
the 4-manifold. Again,
$\tilde{\mathscr{E}}(\mathcal{M})/\mathscr{D}(\mathcal{M})$ is not
a manifold itself, but a stratified space. The points in
$\tilde{\mathscr{E}}(\mathcal{M})/\mathscr{D}(\mathcal{M})$ with
no isometry group form an open and dense subset. In other words,
the stratum of equivalence classes of solutions to Einstein's
equations with no isometries, is open and dense in
$\tilde{\mathscr{E}}(\mathcal{M})/\mathscr{D}(\mathcal{M})$,
(Isenberg and Marsden, p210). This is thought to confirm the
general presumption that space-times with no symmetry are
extremely typical in the set of space-times, and space-times with
some degree of symmetry are extremely special.\footnote{In this
context, the term `special' will be considered equivalent to the
term `atypical'.} The Friedmann-Robertson-Walker models, and
perturbations thereof, considered to be the physically realistic
models for our universe, are believed to be extremely atypical in
the space of all solutions to the Einstein field equations. As
Turner comments, ``even the class of slightly lumpy FRW solutions
occupies only a set of measure zero in the space of initial data,"
(Turner 2001, p655).

This, however, might be a somewhat hasty conclusion to reach. As
Ellis comments: ``We have at present no fully satisfactory measure
of the distance between two cosmological models...or of the
probability of any particular model occurring in the space of all
cosmologies. Without such a solid base, intuitive measures are
often used...the results obtained are dependent on the variables
chosen, and could be misleading-one can change them by changing
the variables used or the associated assumptions. So if one wishes
to talk about the probability of the universe or of specific
cosmological models, as physicists wish to do, the proper
foundation for those concepts is not yet in place," (1999).

When a collection of objects forms a finite-dimensional manifold,
and in the absence of any sort of probability measure, one uses
the Lebesgue measure, or an analogue thereof, to define precisely
what it means for a property to be typical or special. If the
manifold provides a finite measure space, then a property which is
only possessed by a subset of measure zero is considered to be
special, while if the manifold provides an infinite measure space,
then a property which is only possessed by a subset of finite
measure is considered to be special. In both cases, the property
which defines the complement is considered to be typical of the
collection of objects. In the case of an infinite-dimensional
topological vector space, there is no finite, translation
invariant measure to provide an analogue of the Lebesgue measure,
and although non-translation invariant Gaussian measures do exist
on such infinite-dimensional vector spaces, in the case of an
infinite-dimensional manifold there is no diffeomorphism-invariant
measure which is considered to be suitable. As Callender comments,
``debates about likely versus unlikely initial conditions without
a well-defined probability are just intuition-mongering," (2004).
Given the difficulties with finding such a measure, topological
notions of typical and special have been proposed to replace the
measure-theoretic notions.

The first candidate for a topological notion of typicality is the
notion of a dense subset. However, the set of rational numbers is
dense in the set of real numbers, despite having a lower
cardinality, and despite being of Lebesgue measure zero, hence a
dense subset of a topological space is not necessarily considered
to be typical. Instead, following Baire, an open and dense subset
of a topological space is `strongly typical', and a set which
contains the intersection of a countable collection of dense and
open subsets is `residual', or `typical' (Heller 1992, p72). The
irrational numbers are a residual subset of the reals, and
therefore typical, and their complement, the rational numbers, are
atypical. Baire's theorem shows that in a complete metric space or
a locally compact Hausdorff space, a countable intersection of
open dense subsets must itself be dense. However, the attraction
of the Baire definition of typicality is mitigated by the fact
that sets which are open and dense in $\mathbb{R}^n$ can have
arbitrarily small Lebesgue measure, (Hunt \emph{et al} 1992).
Moreover, an infinite-dimensional manifold fails to be locally
compact. It is therefore far from clear that a collection of
universes which is open and dense in a multiverse collection,
should be considered as typical, or that points which belong to
the complement, (those with some degree of symmetry), should be
considered special. There is no \textit{a priori} notion of
typicality on such sets, so it is rather unwise to make such
presumptions. In the classical statistical mechanics of gases, it
is always asserted that a homogeneous distribution of the gas is
the macrostate of highest entropy because it has the greatest
volume of microstates, the greatest volume of phase space,
associated with it. This means that the homogeneous macrostates
must have the highest measure, and certainly not measure zero.
Thus, there is certainly no \textit{a priori} reason from the
finite-dimensional case to think that in the case of continuous
fields, a highly symmetrical configuration should be atypical.

Note that the present universe only approximates a FRW model on
length scales greater than $100$Mpc. On smaller length scales, the
universe exhibits large inhomogeneities and anisotropies. The
distribution of matter is characterised by walls, filaments and
voids up to $100$Mpc, with large peculiar velocities relative to
the rest frame defined by the cosmic microwave background
radiation (CMBR). Whilst the CMBR indicates that the matter in the
universe was spatially isotropic and homogeneous to a high degree
when the universe was $10^{4}-10^{5}$yrs old, the distribution and
motion of galaxies is an indicator of the distribution of matter
in the present era, when the universe is $\sim 10^{10}$ yrs old.
Due to the tendency of gravitation to amplify small initial
inhomogeneities, the level of inhomogeneity in the distribution of
matter has been growing as a function of time.

In contrast with the behaviour of a gas in classical statistical
mechanics, the distribution of matter in a gravitational system
will become more `clumpy' as a function of time. Hence, to
preserve consistency with the second law of thermodynamics it is
generally suggested that for a gravitational system, such clumpy
configurations correspond to higher-entropy macrostates than
configurations with a uniform distribution of matter. Barrow and
Tipler, for example, suggest that a gravitational entropy would
measure the deviation of a universe from exact spatial isotropy
and homogeneity (1986, p446). Penrose suggests that gravitational
entropy is related to the Weyl tensor $C_{\alpha \beta \gamma
\delta}$, which is zero for exact FRW models, but non-zero for the
space-time around a massive body such as a star or black hole.
However, anisotropy does not entail a non-zero Weyl tensor, and a
non-zero Weyl tensor does not entail inhomogeneity. Barrow and
Tipler note that `many' space-times tend towards a plane
gravitational wave geometry, which has a zero Weyl tensor
irrespective of the level of anisotropy (1986, p447). On the other
hand, there are many anisotropic but spatially homogeneous models,
in which the Weyl tensor is non-zero. Thus, if we accept that the
Weyl tensor measures gravitational entropy, then even a universe
which is exactly spatially homogeneous, will, if it exhibits
anisotropic shears and a non-zero Weyl tensor, possess a higher
entropy than a perfectly isotropic and homogeneous FRW universe.

Ellis (2002) suggests that the spatial divergence of the electric
part of the Weyl tensor $E_{\alpha \gamma} = C_{\alpha \beta
\gamma \delta} U^\beta U^\delta$, for a timelike observer vector
field $U$, may be a measure of gravitational entropy. This
quantity is attractive because it does seem to have some
correlation to the level of spatial inhomogeneity. One could take
the integral $\int_\Sigma ||\nabla_a \rho|| \; d\mu$ of the
spacelike gradient of the matter density field $\rho$ as a measure
of the amount of inhomogeneity in the matter field over a
spacelike hypersurface $\Sigma$. In the case of a homogeneous
matter field, this integral vanishes. Ellis suggests that in the
linearized formulation of general relativity,

$$\nabla^a E_{a b} = \frac{1}{3}\nabla_b \rho \;,$$ hence the integral
$\int_\Sigma ||\nabla^a E_{a b}|| \; d \mu$ may, in some
circumstances, be proportional to the level of inhomogeneity, and
thence a good indicator of gravitational entropy. Suppose that one
takes the integral of $\nabla^a E_{a b}$ for an arbitrary
spatially homogeneous but anisotropic model. Being spatially
homogeneous, the spatial gradient of $\rho$ vanishes. If one
introduces an inhomogeneous perturbation to this model, then the
integral of $\nabla^a E_{a b}$ should presumably increase. The
inhomogeneity of our own universe is currently growing, and, if
the inhomogeneous geometries are far more prevalent, in some
sense, than homogeneous geometries, then our universe must be
moving into phase space macrostates of much greater entropy.

There is, however, a problem with the notion that gravitational
entropy should measure the deviation of a universe from exact
isotropy and homogeneity. If there is a positive cosmological
constant, and if Wald's theorem does apply to the case of an
inhomogeneous universe, then the universe will isotropize, and
isotropy requires homogeneity. Even in the absence of a
cosmological constant, all gravitational systems, perhaps even
black holes, will eventually `evaporate', and this might also
entail a return to homogeneity. Thus, if the universe tends
towards either isotropy or homogeneity in the long-term, perhaps
gravitational entropy should not measure the deviation of a
universe from exact isotropy and homogeneity.


\begin{thebibliography}{99}
\bibitem{BarrowTipler86}
Barrow, J.D., Tipler, F.J. (1986). \emph{The anthropic
cosmological principle}, Oxford and NY: Oxford University Press.
\bibitem{Callender}
Callender, C. (2004). Measures, Explanations and the Past: Should
``Special" Initial Conditions Be Explained?, \emph{British Journal
for the Philosophy of Science}, 55(2), pp195-217.
\bibitem{ColHawk73}
Collins, C.B., Hawking, S.W. (1973). Why is the universe
isotropic? \emph{The Astrophysical Journal}, \textbf{180},
pp316-334.
\bibitem{Derdzinski92}
Derdzinski, A. (1992). \emph{Geometry of the Standard Model of
Elementary Particles}, Texts and Monographs in Physics,
Berlin-Heidelberg-New York: Springer Verlag.
\bibitem{Ear99}
Earman, J., and Mosterin, J. (1999). A critical look at
inflationary cosmology, \emph{Philosophy of Science} 66, pp1-49.
\bibitem{Earman02}
Earman, J. (2002). Laws, Symmetry, and Symmetry Breaking;
Invariance, Conservation Principles, and Objectivity, Presidential
address PSA 2002.
\bibitem{Ellis99}
Ellis, G.F.R. (1999). 83 Years of General Relativity and
Cosmology, \emph{Classical and Quantum Gravity}, 16, A37.
\bibitem{Ellis02}
Ellis, G.F.R. (2002). Cosmology and local physics,
\emph{Int.J.Mod.Phys.} A17, pp2667-2672.
\bibitem{End01}
Enderton, H.B. (2001). \emph{A Mathematical Introduction to
Logic}, Second Edition, London: Academic Press.
\bibitem{Hunt92}
Hunt, B.R., Sauer, T., Yorke, J.A. (1992). Prevalence: a
translation-invariant ``almost every'' on infinite-dimensional
spaces. \emph{Bull. Amer. Math. Soc.} 27 pp217-238.
\bibitem{Heller92}
Heller, H. (1992). \emph{Theoretical foundations of cosmology},
Singapore: World Scientific.
\bibitem{Hew97}
Hewitt, C.G., Siklos, S.T.C., Uggla, C., Wainwright, J. (1997).
Exact Bianchi cosmologies and state space, in \emph{Dynamical
systems in cosmology}, eds. J. Wainwright and G.F.R. Ellis,
pp186-211. Cambridge: Cambridge University Press.
\bibitem{IsenMars82}
Isenberg, J., Marsden, J.E. (1982). A Slice Theorem for the Space
of Solutions of Einstein's Equations, \emph{Physics Reports} 89,
No.2, p179-222.
\bibitem{Jant87}
Jantzen, R.T. (1987). Spatially Homogeneous Dynamics: A Unified
Picture, in Proc. Int. Sch. Phys. E. Fermi Course LXXXVI (1982) on
\emph{Gamov Cosmology} (R. Ruffini, F. Melchiorri, Eds.),
pp61-147. Amsterdam: North Holland.
\bibitem{Jen87}
Jensen, L.G., and Stein-Schabes, J.A. (1987). Is inflation
natural?, \emph{Physical Review D35}, pp1146-1150.
\bibitem{Lady98}
Ladyman, J. (1998). What is Structural Realism? \emph{Studies in
History and Philosophy of Science}, 29, pp409-424.
\bibitem{Linde83a}
Linde, A.D. (1983a). Chaotic inflating universe. \emph{Pis'ma v
Zhurnal Eksperimental' noi i Teoreticheskoi Fiziki} 38, pp149-151.
[English translation: Journal of Experimental and Theoretical
Physics Letters 38, pp176-179.]
\bibitem{Linde83b}
Linde, A.D. (1983b). Chaotic inflation. \emph{Physics Letters}
129B, pp177-181.
\bibitem{Mac79}
MacCallum, M.A.H. (1979). Anisotropic and inhomogeneous
relativistic cosmologies, in \emph{General Relativity: An Einstein
Centenary}, eds. S.Hawking and W.Israel, pp533-580. Cambridge:
Cambridge University Press.
\bibitem{Most04}
Mosterin, J. (2004). Anthropic explanations in cosmology, in
\emph{Proceedings of the 12th International Congress of Logic,
Methodology and Philosophy of Science}, Hajek, Valdés and
Westerstahl (eds.). Amsterdam: North-Holland Publishing.
\bibitem{Rucker82}
Rucker, R. (1982). \emph{Infinity and the Mind}. Boston:
Birkhauser.
\bibitem{Smolin97}
Smolin, L. (1997). \emph{The Life of the Cosmos}, London:
Weidenfeld and Nicolson.
\bibitem{Sneed71}
Sneed, J.D. (1971). \emph{The Logical Structure of Mathematical
Physics}, Dordrecht: Reidel.
\bibitem{Suppe89}
Suppe, F. (1989) \emph{The Semantic Conception of Theories and
Scientific Realism}, Urbana, Illinois: University of Illinois
Press.
\bibitem{Suppes69}
Suppes, P. (1969). \emph{Studies in the Methodology and Foundation
of Science: Selected Papers from 1951 to 1969}, Dordrecht: Reidel.
\bibitem{Tegmark98}
Tegmark, M. (1998). Is `the theory of everything' merely the
ultimate ensemble theory?, \emph{Annals of Physics}, 270, pp1-51.
arXiv:gr-qc/9704009.
\bibitem{Turner01}
Turner, M.S. (2001). A Sober Assessment of Cosmology at the New
Millennium. \emph{Publ.Astron.Soc.Pac.} 113 pp653-657
\bibitem{Wald83}
Wald, R.M. (1983). \emph{Phys. Rev. D} 28, 2118.
\bibitem{Wallace01}
Wallace, D. (2001). Worlds in the Everett Interpretation,
\emph{Studies in the History and Philosophy of Modern Physics},
33, pp637-661.
\end{thebibliography}
\end{document}